\documentclass[twocolumn,pra,aps,superscriptaddress, showpacs, amsmath,amssymb,10pt,longbibliography]{revtex4-2} 
\usepackage{amsmath} 
\usepackage{amssymb} 
\usepackage{amsfonts} 
\usepackage{bm} 
\usepackage{amssymb} 
\usepackage{graphicx} 
\usepackage{dcolumn} 
\usepackage{txfonts} 
\usepackage{wasysym} 
\usepackage{makeidx}
\usepackage{color}
\usepackage{mathtools}
\usepackage{threeparttable}
\usepackage[linkcolor=blue,anchorcolor=black,citecolor=blue,colorlinks=true]{hyperref}
\usepackage{breakurl}
\usepackage{float}
\usepackage{soul}
\usepackage{adjustbox}
\usepackage{autobreak}  

\begin{document}
\title{Magnetic Resonance Linewidth of Alkali-Metal Vapor in Unresolved Zeeman Resonance Regime}
\author{Feng Tang}
\affiliation{Beijing Computational Science Research Center, Beijing 100193, PR China}%
\author{Nan Zhao}
\email{nzhao@csrc.ac.cn}
\affiliation{Beijing Computational Science Research Center, Beijing 100193, PR China}%
\date{\today}

\begin{abstract}
The study of magnetic resonance linewidth is crucial in magnetic resonance physics and its applications. 
Previous studies focused on the linewidth of alkali metal atoms within the spin-exchange relaxation-free regime near zero magnetic field 
and in strong magnetic fields where Zeeman resonances are well resolved due to the quadratic Zeeman effect. 
However, the linewidth in the unresolved Zeeman resonance regime, which is prevalent in various magnetometer and comagnetometer applications, is not well understood. 
To address this, we developed a theoretical framework based on the master equation for alkali metal atoms and solved it under the rotating wave approximation and weak driving conditions. 
Our numerical calculations and analytical expressions reveal that the light-narrowing effect occurs only when the ratio of the spin exchange rate to the spin destruction rate exceeds a critical value. 
Additionally, we show that the linewidth in the unresolved Zeeman resonance regime is significantly influenced by the mutual coupling of quantum coherence between different Zeeman sublevels. 
These findings provide a theoretical tool for understanding spin relaxation in alkali-metal atoms and optimizing the performance of atomic magnetometers and comagnetometers operating in this regime.
\end{abstract}

\maketitle

\section{Introduction}
The resonant linewidth is a crucial physical parameter in both fundamental and applied research on magnetic resonance. 
In fundamental studies, the resonant linewidth reveals the interaction between spins and their microscopic environments. 
This linewidth can be utilized to derive various physical insights about the surroundings of the spins, such as the intensity of magnetic field fluctuations and correlation times. 
In practical applications, including magnetic field sensing and magnetic resonance imaging, the magnetic resonance linewidth defines essential performance metrics like magnetic detection sensitivity and resolution. 
Hence, understanding the mechanisms underlying the resonant linewidth is highly important.

Magnetic resonance of spin-polarized alkali metal atoms has been extensively utilized in high-precision measurement domains, 
such as high-sensitivity magnetometers\cite{budkerOpticalMagnetometry2013}, inertial sensors\cite{wrightColdAtomInertial2022,walkerSpinExchangePumpedNMRGyros2016}, and fundamental physical model testing\cite{safronovaSearchNewPhysics2018}. 
The resonant linewidth of alkali-metal atoms primarily depends on the spin-related collision processes involving alkali metal atoms and other atoms or molecules. 
These processes include spin depolarization of alkali metal atoms through spin destruction events and spin exchange processes that lead to spin phase relaxation. 
The spin exchange collision process among alkali metal atoms is crucial in determining the magnetic resonance linewidth. 
For high density alkali metal vapor in low magnetic fields, when the spin-exchange rate $\Gamma_{\mathrm{SE}}$ considerably exceeds the Larmor precession frequency of spin rotation $\Omega_{\mathrm{L}}$ (i.e., $\Gamma_{\mathrm{SE}} \gg \Omega_{\mathrm{L}}$), the system enters a spin-exchange relaxation-free regime (SERF) with a notable reduction in the resonant linewidth \cite{happerEffectRapidSpin1977}. 
Increasing the magnetic field violates the SERF conditions, causing the spin resonance linewidth of high density alkali metal vapor to be primarily determined by the spin exchange rate. 
In this situation, the resonant linewidth also depends on the level of atomic spin polarization, and increasing atomic spin polarization via optical pumping results in the light-narrowing effect \cite{bhaskarLightNarrowingMagnetic1981,appeltLightNarrowingRubidium1999}. 
A quantitative understanding of the various collision-induced physical effects is essential for the use of alkali metal vapor in precision measurements.

The master equation is used to describe various microscopic physical processes that influence the spin resonance linewidth \cite{Appelt1998}.
The microscopic processes can be quantitatively characterized by their collision cross-sections, and magnetic resonance linewidth in obtained by solving the master equation quantitatively under different conditions. 
In Ref.~\cite{Appelt1998}, Appelt et al. extensively analysed the spin relaxation mechanisms of alkali metal atoms and calculated the linewidth within the resolved Zeeman resonance (RZR) regime, 
where the frequency degeneracy of coherences between adjacent Zeeman levels is lifted by the quadratic Zeeman effect in strong fields \cite{schiffTheoryQuadraticZeeman1939,julsgaardCharacterizingSpinState2004}. 
Unfortunately, because of the complexity of alkali metal energy levels and the nonlinear nature of the spin-exchange process, 
calculating the magnetic resonance linewidth under more general conditions remains a challenging task.

Magnetic resonance linewidths show varying characteristics depending on the strength of the magnetic field and the spin polarization. 
Extensive studies have been conducted on resonance linewidths under both weak and strong magnetic field conditions. 
In the SERF regime at near-zero magnetic fields, the resonance linewidth is inversely proportional to the spin exchange rate\cite{happerEffectRapidSpin1977}. 
In strong magnetic fields, the behavior of the magnetic resonance linewidth in the RZR regime has been explored in Refs.\cite{Appelt1998, appeltLightNarrowingRubidium1999, jauIntenseNarrowAtomicClock2004}. 
Conversely, in the intermediate magnetic field region, the interaction between quantum coherences of different Zeeman sublevels plays a significant role in influencing the magnetic resonance linewidth. 
A precise description of the linewidth in the unresolved Zeeman resonance (UZR) regime remains unavailable. The advance in various atomic magnetometers operating in the UZR regime necessitates a more thorough examination of the linewidth.

Based on the master equation in Ref.~\cite{Appelt1998}, we investigated the linewidth of the atom resonance under various conditions. 
We analyze the mathematical structure of the evolution matrix of the master equation, simplifying it under the rotating wave approximation and weak driving scenarios. 
This simplification not only makes the master equation more tractable, but also offers a clearer understanding of the physical phenomena under differing conditions. 
Our findings indicate that, in the UZR regime, the interplay of quantum coherence among various Zeeman sublevels has a significant impact on the magnetic resonance linewidth. 
We derived analytical expressions for the linewidth in both high- and low-spin polarization limits, which will benefit the optimization of parameters for atomic magnetometers and gyroscopes operating in the UZR regime.

\section{Theoretical Treatment}
\begin{figure*}[tbp]  
    \includegraphics[width=0.6\textwidth]{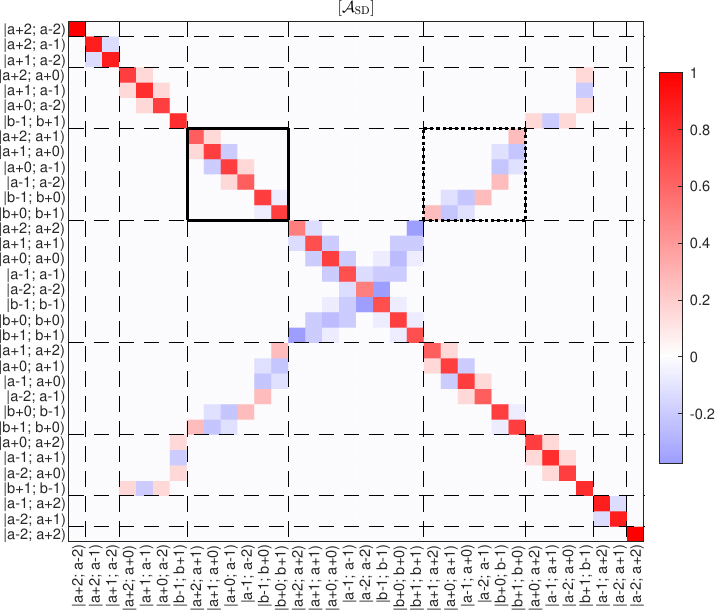}  
    \caption{Heat map plot of the matrix $\left[\mathcal{A}_{\rm SD}\right]$. The bases are denoted by $\vert F, m_F; F', m'_{F'}) = \left\vert m_F; m'_{F'}\right)$ [see Eq.~\eqref{eq:basis}], 
    and are grouped according to their frequencies in the rotating frame. 
    The horizontal and vertical dashed lines separate the matrix blocks according to their frequencies in the rotating frame.
    The block marked by the solid lines is $\left[\mathcal{A}_{\rm SD}\right]_{1, 1}$, the diagonal block of the first-order Zeeman coherences.
    The block marked by the dotted lines is $\left[\mathcal{A}_{\rm SD}\right]_{1, -1}$, which is a double frequency block in the rotating frame and is neglected with the RWA.}
    \label{fig:matrix}  
\end{figure*}

\subsection{Master equation in Liouville space}
We study alkali metal spins in a static magnetic field $B_0$ in the $\hat{z}$ direction and a radio-frequency (RF) driving field in the $\hat{x}$ direction. 
An circularly polarized laser beam travels along the $\hat{z}$ axis to pump the alkali metal atoms. 
The primary incoherent interactions that affect the spin dynamics of alkali metals are spin destruction collisions between alkali metal atoms and other atoms and molecules and spin exchange collisions among alkali metal atoms. 
The master equation governing the density matrix $\rho$ of the alkali-metal atom is \cite{Appelt1998}
\begin{align}\label{MasterEq1}
\frac{d\rho}{dt} =& -i [H,\rho]+ \Gamma_{\text{SE}}[\varphi(1+4\langle \mathbf{S} \rangle\cdot \mathbf{S}) -\rho]  \nonumber \\
&+ \Gamma_{\text{SD}}(\varphi-\rho) + R_{\text{OP}} [\varphi(1+2 S_z)-\rho] .
\end{align}
The coherent evolution is determined by the Hamiltonian 
\begin{equation}
    H = H_0 + H_{\text{drv}},
\end{equation}
where 
\begin{align}
H_0 &= \Omega_0 S_z, \label{eq:hyperfine} + \Omega_{\rm hf} \mathbf{I} \cdot \mathbf{S} \\
H_{\text{drv}} &=\Omega_R S_x \cos(\omega t + \phi_0),\label{eq:rabi}
\end{align}
represent the atomic spin Hamiltonian within the magnetic field $B_0$, along with the RF driving field. 
In Eq.~\eqref{eq:hyperfine}, $\mathbf{S}$ and $\mathbf{I}$ correspond to the electronic spin and nuclear spin operators of the alkali metal atom. 
Here, $\Omega_0 =\vert \gamma_e B_0 \vert > 0$ denotes the Larmor frequency ($\gamma_e$ being the gyromagnetic ratio), and $\Omega_{\rm hf}$ is the hyperfine interaction constant. 
Throughout this paper, it is assumed that the magnetic field $B_0$ is sufficiently weak so that $\Omega_0 \ll \Omega_{\rm hf}$ holds true. 
In Eq.~(\ref{eq:rabi}), $ \Omega_R $, $\omega$ and $\phi_0$ stand for the Rabi frequency, the frequency and the phase of the RF driving field, respectively. 
For simplicity, $\phi_0 = 0$ is assumed in the following text.

The incoherent part of the master Eq.~\eqref{MasterEq1} comprises spin exchange (SE) collisions, spin destruction (SD) collisions, and optical pumping (OP) mechanisms. 
The second term in Eq.~\eqref{MasterEq1} characterizes the spin exchange collisions among alkali metal spins at a rate of $\Gamma_{\text{SE}}$, 
where $\langle \mathbf{S}\rangle = \text{tr}[\rho \mathbf{S}]$ represents the spin expectation value.
The third term results from spin destruction collisions between alkali metal atoms and other atoms or molecules, 
with a total spin destruction rate denoted as $\Gamma_{\text{SD}}$.
In Eq.~\eqref{MasterEq1}, we have assumed that the optical pumping occurs in the high broadening limit of the optical transition\cite{happerOpticalPumping1972, happerOpticallyPumpedAtoms2010},
and the pumping effect is described by the final term of Eq.~\eqref{MasterEq1} with a pumping rate $R_{\text{OP}}$. 
The state 
\begin{equation}
    \varphi = \frac{1}{4}\rho  + \mathbf{S}\cdot \rho \mathbf{S}
\end{equation}
is a purely nuclear spin operator of the alkali-metal atom, without electronic spin polarization \cite{happerOpticalPumping1972}.

In Liouville space, the density matrix $\rho$ is mapped to a state vector $|\rho)$. 
Thee state vector is expanded as
\begin{equation}
    |\rho) = \sum_{m,m',F,F'}  |m_F;m'_{F'})(m_F;m'_{F'}|\rho ),
\end{equation}
where 
\begin{equation}
\vert m_F; m'_{F'}) = \vert F, m\rangle \langle F', m'\vert
\label{eq:basis}
\end{equation}
is the basis vector of the Liouville space, 
and $\vert F, m\rangle$ is the eigen vector of the alkali-metal Hamiltonian \eqref{eq:hyperfine}, 
with the total angular momentum quantum number $F=I+1/2\equiv a$ or $F=I-1/2\equiv b$,
and the magnetic quantum number $m=-F, \dots, F$.
For the state vectors constrained in the subspace of population and Zeeman coherences with $F=F'$, we define
\begin{equation}
    \vert F, \bar{m})_{\Delta m} = \vert m_F; m'_F),
\end{equation}
where $\bar{m}= (m+m')/2$ and $\Delta m = m-m'$ \cite{Appelt1998}.

The master Eq.~\eqref{MasterEq1} in Liouville space becomes
\begin{align}\label{MasterEqL}
\frac{d}{dt} |\rho) =& - i \mathcal{H} |\rho) - (\Gamma_{\text{SE}} + \Gamma_{\text{SD}} + R_{\text{OP}}) \mathcal{A}_{\text{SD}}|\rho)\nonumber\\
& + \left(\Gamma_{\text{SE}} ( \mathbf{S}\vert \rho ) +  \frac{1}{2}R_{\rm op}\hat{z} \right) \cdot \boldsymbol{\mathcal{A}}_{\text{SE}} |\rho) \equiv  -\mathcal{G} |\rho),
\end{align}
where $\langle \mathbf{S}\rangle  = ( \mathbf{S}\vert \rho )$ is the spin expectation value expressed in Liouville space,
and $\mathcal{H}$, $\mathcal{A}_{\rm SD}$ and  $\boldsymbol{\mathcal{A}}_{\rm SE}$ are the superoperator in the Liouville space, corresponding to the coherent evolution, spin destruction and spin exchange processes, respectively.
The superoperators are defined as \cite{happerOpticallyPumpedAtoms2010}
\begin{align}
    \mathcal{H} &= \mathcal{H}_0 +\mathcal{H}_{\rm drv}  =   H^{\copyright}_0  +  H_{\rm drv}^{\copyright},\\
    \mathcal{A}_{\text{SD}} &= \frac{1}{2} \mathbf{S}^{\copyright} \cdot \mathbf{S}^{\copyright},\\
    \boldsymbol{\mathcal{A}}_{\text{SE}} &= \mathbf{S}^{\flat} + \mathbf{S}^{\sharp} - 2 i \mathbf{S}^{\flat} \times \mathbf{S}^{\sharp},
\end{align}
where $X^{\flat}$ and $X^{\sharp}$ are the left- and right-translation superoperators, corresponding to the operator $X$ in Hilbert space, and $X^{\copyright} = X^{\flat} - X^{\sharp}$ is the commutator superoperator \cite{happerOpticallyPumpedAtoms2010}.

\subsection{Rotating wave approximation}
We consider the near-resonant RF driving field with frequency $\omega \approx \Omega_0$. 
The driving field only induces coherences between the Zeeman sublevels within the $F=a$ or $b$ subspaces. 
In the absence of microwave excitations, the hyperfine coherences between levels $F=a$ and $F=b$ (e.g., $\vert m_a; m'_b)$) will decay within the time scale of the hyperfine relaxation time, typically in the order of $\sim {\rm ms}$ or even shorter.
In this case, we can neglect the hyperfine coherences and focus on the dynamical evolution of the population and Zeeman coherences.
To this end, we define the projection operators
\begin{align}
    \mathcal{P}^{(\text{hf})}_a &=  \sum_{m, n} \vert m_a; n_a)(m_a; n_a\vert, \\
    \mathcal{P}^{(\text{hf})}_b &=  \sum_{m, n} \vert m_b; n_b)(m_b; n_b\vert, \\
    \mathcal{P}^{(\text{hf})}_{\text{Z}} &= \mathcal{P}^{(\text{hf})}_a + \mathcal{P}^{(\text{hf})}_b,
\end{align}
and the state vector projected to the population and Zeeman coherence subspace is
\begin{equation}
|\rho_{\text{Z}}) = \mathcal{P}_{\text{Z}}^{(\text{hf})}|\rho). 
\end{equation}
The evolution of $|\rho_{\text{Z}})$ is governed by
\begin{align} \label{Materhf_RWA0}
\frac{d}{dt} |\rho_{\text{Z}}) = - \mathcal{P}_{\text{Z}}^{(\text{hf})} \mathcal{G} \mathcal{P}_{\text{Z}}^{(\text{hf})}  \vert \rho_{\text{Z}}) \equiv -\mathcal{G}_{{\text{Z}}} |\rho_{\text{Z}})
\end{align}
where 
\begin{align} \label{Ghf_RWA}
    \mathcal{G}_{{\text{Z}}} = i\left[\mathcal{H}\right] + \Gamma_{\text{tot}} \left[\mathcal{A}_{\text{SD}}\right]  -\left(\Gamma_{\text{SE}}   (\mathbf{S}|\rho_Z) + \frac{1}{2} R_{\text{OP}}\hat{z} \right) \cdot \left[\boldsymbol{\mathcal{A}}_{\text{SE}}\right],
\end{align}
where $\Gamma_{\text{tot}}=\Gamma_{\text{SE}} + \Gamma_{\text{SD}} +R_{\text{OP}}$, 
and $\left[ \mathcal{X}\right] = \mathcal{P}^{(\text{hf})}_{\text{Z}} \mathcal{X}\mathcal{P}^{(\text{hf})}_{\text{Z}}$ represents the projection of the superoperator $\mathcal{X}$ to the subspace of population and  Zeeman coherences.
As an example, Fig.~\ref{fig:matrix} shows the matrix of the superoperator $\mathcal{A}_{\rm SD}$ on the basis of the population and Zeeman coherences.

To further simplify the master equation, we define the rotating frame with the frequency $\omega$ and make the rotating wave approximation (RWA).
We introduce the $k$th order Zeeman coherence projector.
\begin{align}
    \mathcal{P}_k = \sum_{\bar{m}} |a,\bar{m})_{k} ~_{k}(a,\bar{m}| + \sum_{\bar{m}} |b,\bar{m})_{-k} ~_{-k}(b,\bar{m}|,
\end{align}
and the free evolution superoperator
\begin{align}
\mathcal{G}_0 = i \sum_k k\omega \mathcal{P}_k.
\end{align}
With the rotation transform generated by $\mathcal{G}_0$, i.e., 
\begin{align}
|\rho_{\text{Z}}) &= \exp\left( -\mathcal{G}_0  t \right) |\tilde{\rho}_{\text{Z}}) \equiv \mathcal{U}(t) |\tilde{\rho}),
\end{align}
the master equation of the slow-varying state vector $|\tilde{\rho})$ in the rotating frame is
\begin{align}
\frac{d}{dt}|\tilde{\rho}) = -\left( \mathcal{U}^{\dagger} \mathcal{G}_{\text{Z}} \mathcal{U} - \mathcal{G}_0\right) \vert \tilde{\rho}) \equiv  -\mathcal{G}_{\text{rot}}(t) |\tilde{\rho}).
\end{align}
The evolution superoperator $\mathcal{G}_{\text{rot}}(t)$  is
\begin{equation}
    \label{eq:Gdrv_general}
    \mathcal{G}_{\text{rot}}(t) = i \sum_k \left[\Delta\right]_{k,k}+ \mathcal{G}_{\text{drv}}(t) + \mathcal{G}_{\text{relax}}(t),  
\end{equation}
where $\left[\Delta\right]_{k, k} \equiv \mathcal{P}_k \mathcal{H}_0 \mathcal{P}_k  -k\omega \mathcal{I}_k$ is a diagonal matrix representing the frequency detuning of the $k$th order coherences ($\mathcal{I}_k$ is an identity matrix).
The driving superoperator $\mathcal{G}_{\text{drv}}(t)$ in the rotating frame is
\begin{equation}
    \label{eq:Gdrv_Rotating}
    \mathcal{G}_{\text{drv}}(t) =  i \Omega_R\cos(\omega t) \sum_{p,q} e^{i(p-q)\omega t} \left[S_x^{\copyright}\right]_{p,q} ,
\end{equation}
and the relaxation superoperator $\mathcal{G}_{\text{relax}}(t)$ is
\begin{align}
    \label{eq:Grelax_Rotating}
    &\mathcal{G}_{\text{relax}}(t)  = \Gamma_{\text{tot}} \sum_{p,q} e^{i(p-q)\omega t} \left[\mathcal{A}_{\text{SD}}\right]_{p,q} \nonumber \\
     &-\left(\Gamma_{\text{SE}}  \sum_k  \tilde{\mathbf{S}}^{(k)} e^{-i k \omega t}+ \frac{1}{2}R_{\text{OP}} \hat{z}\right) \cdot 
     \sum_{p,q} e^{i(p-q)\omega t}\left[\boldsymbol{\mathcal{A}}_{\text{SE}}\right]_{p,q}.
\end{align}
In Eqs.~\eqref{eq:Gdrv_general} - \eqref{eq:Grelax_Rotating}, $\left[ \mathcal{X}\right]_{p,q} = \mathcal{P}_p \left[\mathcal{X}\right]\mathcal{P}_q$ is the block matrix of the $p$th row and the $q$th column,
corresponding to a time-dependent factor $\exp(i(p-q)\omega t)$ in the rotating frame.
The spin expectation value in the rotating frame is
\begin{align}
   \langle \mathbf{S}\rangle &= (\mathbf{S}\vert \rho_{\text{Z}}) = (\mathbf{S}\vert e^{-\mathcal{G}_{0}t}\vert \tilde{\rho}) \nonumber\\
   &= \sum_{k} (\mathbf{S}\vert \mathcal{P}_{k}\vert \tilde{\rho})e^{-ik \omega t} \equiv \sum_{k} \tilde{\mathbf{S}}^{(k)} e^{-ik\omega t},
\end{align}
where 
\begin{equation}
    \label{eq:S_rotating_frame}
    \tilde{\mathbf{S}}^{(k)} = (\mathbf{S}\vert \mathcal{P}_{k}\vert \tilde{\rho}) = (\mathbf{S}\vert \tilde{\rho}_k)
\end{equation}
 is the spin expectation value on the $k$th order Zeeman coherence $\vert \tilde{\rho}_k) = \mathcal{P}_k \vert \tilde{\rho})$.

Keeping only the zero-frequency terms of $ \mathcal{G}_{\text{rot}}(t)$, we obtain the RWA equation
\begin{equation}
    \label{eq:RWA_master_equation}
    \frac{d}{dt} |\tilde{\rho}) = - \mathcal{G}_{\text{RWA}}|\tilde{\rho}),
\end{equation}
where the evolution superoperator $\mathcal{G}_{\text{RWA}}$ can be decomposed as $\mathcal{G}_{\text{RWA}}= \sum_k \mathcal{G}^{(k)}_{\text{RWA}}$ with 
\begin{align} \label{MasterZeeman_Rho}
    \mathcal{G}^{(k)}_{\text{RWA}} &= i \left[\Delta\right]_{k, k} + i \frac{ \Omega_R}{2} \left( [S_x^{\copyright}]_{k,k+1} +   [S_x^{\copyright}]_{k,k-1}   \right) \nonumber \\
    & +   (\Gamma_{\text{SE}}+\Gamma_{\text{SD}}+R_{\text{OP}})  \left[\mathcal{A}_{\text{SD}}\right]_{k,k}- \frac{R_{\text{OP}}}{2}   \left[\mathcal{A}_{\text{SE}}^{(z)}\right]_{k,k} \nonumber \\
    &-\Gamma_{\text{SE}}\left(   \mathcal{A}_{\text{SE}}^{(k,z)}+\mathcal{A}_{\text{SE}}^{(k, +)} +\mathcal{A}_{\text{SE}}^{(k, -)}\right).
\end{align}
In Eq.~\eqref{MasterZeeman_Rho}, the spin exchange due to the transverse and longitudinal spin components are
\begin{align}
    \mathcal{A}_{\text{SE}}^{(k, \pm)} &= \tilde{S}_x^{(\mp 1)} \left[\mathcal{A}_{\text{SE}}^{(x)}\right]_{k,k\pm 1} + \tilde{S}_y^{(\mp 1)} \left[\mathcal{A}_{\text{SE}}^{(y)}\right]_{k,k\pm 1} 
\end{align}
and 
\begin{align}
    \mathcal{A}_{\text{SE}}^{(k, z)} &=  \tilde{S}_z^{(0)} \left[\mathcal{A}_{\text{SE}}^{(z)}\right]_{k,k}  .
\end{align}

\begin{figure}[tbp]  
    \centering  
    \includegraphics[width=0.9\columnwidth]{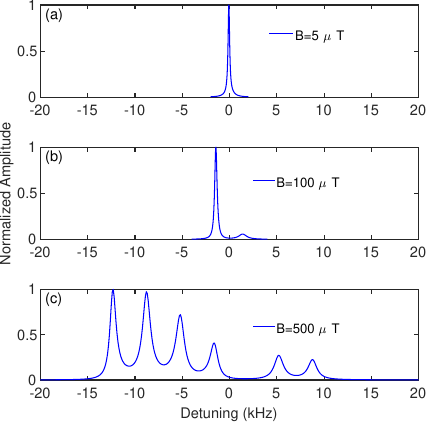}  
    \caption{(a) Magnetic resonance spectrum of weakly polarized (i.e., $P\ll 1$) ${}^{87}{\rm Rb}$ atoms in a weak field of $5~{\rm \mu T}$.
      (b) The same as (a), but for a stronger field of $100~{\rm \mu T}$. Two resonance peaks corresponds to the resonances of $F=2$ and $F=1$ levels, respectively.
      (c) The same as (a), but for a field of $500~{\rm \mu T}$, which drives the atom in the RZR regime. 
      Six resonance peaks are well resolved, and the peaks are broader than the weak field cases.
      The parameters used in the calculation are $\Gamma_{\rm SD} = 10~{\rm Hz}$ and $\Gamma_{\rm SE} = 775~{\rm Hz}$.
      A very weak optical pumping rate $R_{\rm op} =  1~{\rm Hz}$ is assumed to ensure the weak spin polarization.}
    \label{fig0:spectrum}  
\end{figure}

\subsection{Resolved and unresolved Zeeman resonance regimes}
With the RWA, the evolution superoperator $\mathcal{G}_{\rm RWA}$ is time-independent. 
The magnetic resonance spectrum is obtained from the steady-state solution of Eq.~\eqref{eq:RWA_master_equation}. 
Figure~\ref{fig0:spectrum} shows the numerical results of the spectrum of weakly polarized ${}^{87}{\rm Rb}$ atoms in different magnetic fields.

In a weak field, the frequency difference between the $k$th order Zeeman coherences is negligible, 
and the detuning matrix $\left[\Delta\right]_{k, k}$ in Eq.~\eqref{MasterZeeman_Rho} is proportional to an identity matrix, i.e., $\left[\Delta\right]_{k, k} = \Delta_k \mathcal{I}_k$.
In this case, all Zeeman resonances are degenerate, and only one resonance peak appears in the spectrum [see Fig.~\ref{fig0:spectrum}(a)].

As the magnetic field increases, the frequency degeneracy of Zeeman coherences in subspaces with $F=a$ and $F=b$ is lifted because of the slight difference of their gyromagnetic ratios.
Taking ${}^{87}{\rm Rb}$ atoms as an example, the difference between the gyromagnetic ratios of $F=2$ and $F=1$ levels is $\Delta \gamma =\vert \gamma_b\vert -\vert \gamma_a\vert = 2\pi \times  28~{\rm Hz/\mu T}$,
which causes a splitting of $2\pi \times 2.8~{\rm kHz}$ in a field of $100~{\rm \mu T}$ [see Fig.~\ref{fig0:spectrum}(b)].
In this field, the Zeeman coherences within the $F=1$ or $F=2$ subspaces can still be regarded as degenerate, and the spectrum consists of two peaks.

When the magnetic field is so strong that the quadratic splitting $\omega_{Q} = 2(\gamma_{e} B )^2/ \Omega_{\rm hf}$ well exceeds the typical resonance linewidth \cite{schiffTheoryQuadraticZeeman1939,julsgaardCharacterizingSpinState2004},
the system enters the RZR regime. 
Figure~\ref{fig0:spectrum}(c) shows the example of ${}^{87}{\rm Rb}$ atoms in a magnetic field $B_0=500~{\rm \mu T}$, 
where the spectrum consists of six peaks with quadratic splitting $\omega_Q = 2\pi \times 3.6~{\rm kHz}$.

As shown in Fig.~\ref{fig0:spectrum}, the resonance linewidth in the RZR regime is much broader than that in the UZR regime.
The linewidth in the RZR regime [Fig.~\ref{fig0:spectrum}(c)] has been well studied theoretically \cite{Appelt1998} and experimentally \cite{appeltLightNarrowingRubidium1999}.
In the following, we will focus on the linewidth in the UZR regime.
\subsection{Weak driving approximation}
\subsubsection{Weak driving approximation in UZR regime}

With the operator $\mathcal{P}_1$ applied on both sides of Eqs.~\eqref{eq:RWA_master_equation}, 
the master equation satisfied by the first order Zeeman coherence $|\tilde{\rho}_1)$ is
\begin{align} \label{FirstOrder1}
\frac{d}{dt} |\tilde{\rho}_1) =& - \left[\mathcal{G}_{\text{RWA}}\right]_{1, 1} \vert\tilde{\rho}_1) - \left[\mathcal{G}_{\text{RWA}}\right]_{1, 0} \vert\tilde{\rho}_0) - \left[\mathcal{G}_{\text{RWA}}\right]_{1, 2} \vert\tilde{\rho}_2),
\end{align}
where the diagonal block is
\begin{align} 
    \label{FirstOrder1_G}
\left[\mathcal{G}_{\text{RWA}}\right]_{1, 1} =& i \Delta_1 +\Gamma_{\text{tot}} \left[\mathcal{A}_{\text{SD}}\right]_{1,1} -\left( \frac{R_{\text{OP}}}{2}   + \Gamma_{\text{SE}}  \tilde{S}_z^{(0)} \right) \left[\mathcal{A}^{(z)}_{\text{SE}}\right]_{1,1}, 
\end{align}
and the off-diagonal blocks 
\begin{align}
\left[\mathcal{G}_{\text{RWA}}\right]_{1,0} =&  \frac{i \Omega_R}{2} \left[S_x^{\copyright}\right]_{1,0} - \Gamma_{\text{SE} } \mathcal{A}_{\text{SE}}^{(k,-)}, \\
\left[\mathcal{G}_{\text{RWA}}\right]_{1,2} =&  \frac{i \Omega_R}{2} \left[S_x^{\copyright}\right]_{1,2} - \Gamma_{\text{SE} } \mathcal{A}_{\text{SE}}^{(k,+)},
\end{align}
couple the first order Zeeman coherence $|\tilde{\rho}_1)$ to the population $\vert \tilde{\rho}_0)$ and the second order Zeeman coherence $|\tilde{\rho}_2)$, due to the RF driving field and the transverse components of the spin exchange.
In Eq.~\eqref{FirstOrder1_G}, we have replaced the detuning matrix $\left[\Delta\right]_{1, 1}$ by the detuning frequency $\Delta_1$ (the identity matrix is omitted for simplicity) in the UZR regime.

In a weak RF driving field, the $k$th order Zeeman coherence $|\tilde{\rho}_{k})$ is on the order of $(\Omega_{R}/\Gamma_2)^k$ \cite{Appelt1998}, where $\Gamma_2$ is the resonance linewidth.
Although the exact value of $\Gamma_2$ is unknown before solving the master equation, it is reasonable that $\Gamma_2$ is in the same order as the relaxation rates $\Gamma_2 \sim \Gamma_{\text{SE}}$, $\Gamma_{\text{SD}}$ or $R_{\text{OP}}$.
We can always control the strength of the RF driving field, so that $\Omega_R \ll \Gamma_2$ is well satisfied. 
In this case, the population $|\tilde{\rho}_0)$ is approximately constant, and we replace $|\tilde{\rho}_0)$ in Eq.~\eqref{FirstOrder1} by the equilibrium state $\vert \rho_{\text{eq}})$ in the absence of RF driving (with optical pumping and spin relaxation only), i.e.,
\begin{equation}
|\tilde{\rho}_0) = \mathcal{P}_0 \vert \rho_{\text{eq}}) \equiv |\bar{\rho}_0).
\end{equation} 
Furthermore, the second order Zeeman coherence $|\tilde{\rho}_2)$ in Eq.~\eqref{FirstOrder1} is neglected.
With this weak driving approximation (WDA) Eq.~\eqref{FirstOrder1} becomes
\begin{align}
    \label{FirstOrder1WDA}
    \frac{d}{dt} |\tilde{\rho}_1) &= - \left( \left[\mathcal{G}_{\text{RWA}}\right]_{1, 1} - \Gamma_{\rm SE} \left[\mathcal{A}_{\text{SE} }^{(\perp)}\right]_{1,1} \right)  \vert \tilde{\rho}_1) -  \Omega_R \vert \bar{\nu}_1 ) \nonumber\\
    & \equiv - \mathcal{G}_{\text{WDA}} \vert \tilde{\rho}_1 ) -  \Omega_R \vert \bar{\nu}_1)
\end{align}
In Eq.~\eqref{FirstOrder1WDA}, the superoperator due to the spin exchange of transverse components is 
\begin{align}
    \left[\mathcal{A}_{\text{SE} }^{(\perp)}\right]_{1, 1} &= \left[\mathcal{A}_{\text{SE}}^{(x)}\right]_{1,0} \left[\mathcal{Q}_{x}\right]_{0,1} + \left[\mathcal{A}_{\text{SE}}^{(y)}\right]_{1,0} \left[\mathcal{Q}_{y}\right]_{0,1}, \label{GWDA}
\end{align}
where $\left[\mathcal{Q}_{x}\right]_{0, 1}$ and $\left[\mathcal{Q}_y\right]_{0, 1}$ are matrices associated with the equilibrium state $\vert \rho_{\text{eq}})$ as
\begin{equation}
    \left[\mathcal{Q}_{x,y}\right]_{0, 1} = \mathcal{P}_0 \vert \rho_{\text{eq}}) (S_{x, y}\vert \mathcal{P}_1.\label{Qxy}
\end{equation}
The inhomogeneous term of Eq.~\eqref{FirstOrder1WDA} arises from the RF driving field, with the state vector $\vert \bar{\nu}_1)$ given by
\begin{equation}
    \vert \bar{\nu}_1) = \frac{i}{2} \left[S_x^{\copyright}\right]_{1, 0} \vert \bar{\rho}_0). \label{nu1}
\end{equation}
Notice that, with the WDA, Eq.~\eqref{FirstOrder1WDA} is a closed linear equation for $|\tilde{\rho}_1)$ and is used to analyze the resonance linewidth.

\subsubsection{Spin temperature distribution}
 In general, the WDA does not have requirement of the specific form of the population $\vert \bar{\rho}_0)$. 
 Here we consider the spin temperature distribution, which is widely used in the study of alkali-metal spin dynamics \cite{Appelt1998}.
 The spin temperature distribution density matrix is 
 \begin{align}
\rho_{\rm eq} = \frac{e^{\beta F_z}}{Z} = \frac{e^{\beta I_z}e^{\beta S_z} }{Z_I Z_S},
 \end{align}
where $F_z=S_z+I_z$, $Z_J = \sum_{m=-J}^{J} \exp(\beta m)$ and $Z=Z_{I}Z_{S}={\rm Tr}[\exp(\beta F_z)]$ is the normalization factor. 
The parameter $\beta$ is related to the electron spin polarization $P$ by
\begin{align}
    P = 2 \langle S_z \rangle = 2 \tilde{S}_z^{(0)} = \tanh\left(\frac{\beta}{2}\right). 
\end{align}
The probability of occupation $p_m$ of the spin state $\vert F, m\rangle $ is a function of spin polarization $P$
\begin{align}
p_m(P) =  \frac{P(1+P)^{[I]/2+m}(1-P)^{[I]/2-m} }{(1+P)^{[I]}-(1-P)^{[I]}},
\end{align}
where $[I]=2I+1$.
As the state $\rho_{\rm eq} = \rho_{\rm eq}(P)$ is determined by $P$, the matrices $\mathcal{Q}_{x, y} = \mathcal{Q}_{x,y}(P)$ defined in Eq.~\eqref{Qxy}, 
together with the superoperator $\left[\mathcal{A}_{\text{SE} }^{(\perp)}\right]_{1, 1}=\left[\mathcal{A}_{\text{SE}}^{(\perp)}(P)\right]_{1, 1}$ defined in Eq.~\eqref{GWDA}, are also functions of $P$.

With the spin temperature distribution, the dimensionless WDA superoperator $\mathbb{G}_{\text{WDA}}$ (in the unit of $\Gamma_{\rm SD}$) becomes
\begin{align}
    \label{eq:dimensionless_GWDA}
  &\mathbb{G}_{\rm WDA}(\eta, P) \equiv \frac{\mathcal{G}_\text{WDA}}{\Gamma_{\rm SD}}  \\
  =&  \left( \eta+ \frac{1}{1-P}\right) \left(\left[\mathcal{A}_{\text{SD}}\right]_{1,1}  -   \frac{P}{2}\left[\mathcal{A}_{\text{SE}}^{(z)}\right]_{1,1}\right) - \eta \left[\mathcal{A}_{\text{SE}}^{(\perp)}(P)\right]_{1, 1},\nonumber
\end{align}
where $\eta= \Gamma_{\rm SE}/\Gamma_{\rm SD}$.
In Eq.~\eqref{eq:dimensionless_GWDA}, we have used the fact that the spin polarization $P$ of the spin temperature state is 
\begin{eqnarray}
    P = \frac{R_{\rm op}}{R_{\rm op}  +  \Gamma_{\rm SD}},
\end{eqnarray}
or, equivalently, the dimensionless pumping rate is expressed in terms of $P$ as
\begin{eqnarray}
    \frac{R_{\rm op}}{\Gamma_{\rm SD}} = \frac{P}{1-P}.
\end{eqnarray}
Notice that the dimensionless WDA superoperator $\mathbb{G}_{\text{WDA}}(\eta, P)$ is fully characterized by two dimensionless parameters, namely, 
the relative strength $\eta$ of spin exchange and the spin polarization $P$.
  
\subsubsection{Observable and lineshape}
In typical magnetometer experiments, a linearly polarized laser beam is used to measure the transverse spin component, e.g. $S_x$, via the Faraday rotation effect. 
According to the discussion above, the time-dependent observable $\langle S_x(t)\rangle$ is 
\begin{align} \label{Sx_Average}
\langle S_x (t)\rangle = (S_x|\rho_{\rm Z}(t)) = \tilde{S}_x^{(1)} e^{-i \omega t} + \text{c.c},
\end{align}
and the in-phase and out-of-phase quadratures are 
\begin{align}
    X &= \text{Re} \left[  \tilde{S}_x^{(1)}  \right] = \text{Re}\left[ (S_x|\tilde{\rho}_1)  \right],\\
    Y &=  -\text{Im} \left[  \tilde{S}_x^{(1)}  \right]= -\text{Im}\left[ (S_x|\tilde{\rho}_1)  \right]. 
\end{align}

\begin{figure}[tbp]  
    \centering  
    \includegraphics[width=\columnwidth]{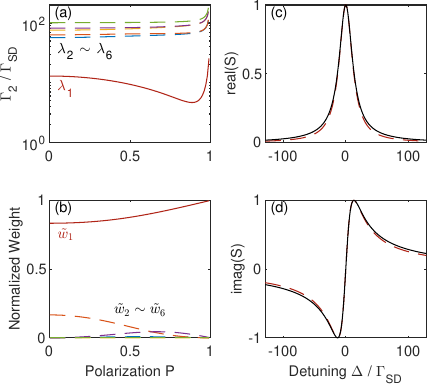}  
    \caption{(a) The eigenvalues of $\mathbb{G}_{\rm{WDA}}$ as functions of $P$ with $\eta=100$. The lowest curve represents the smallest eigenvalue $\lambda_1$, while the dashed curves are $\lambda_2, \lambda_3, \dots, \lambda_6$.
     (b) The normalized weight $\tilde{w}_k = \vert w_k\vert / \sum_i \vert w_i\vert$. The red solid curve is the weight corresponding to $\lambda_1$. Dashed curves are much smaller weights corresponding to other eigenvalues.
     (c) and (d) The real and imaginary part of the observable $S$, normalized to the maximum value. The solid curve is the sum of contributions of all eigenvalues [see Eq.~\eqref{eq:observable}], 
     while the dashed curve is the only contribution corresponding to the smallest eigenvalue $\lambda_1$.}
    \label{EigenvaluesOfG}  
\end{figure}

\begin{figure*}[thb]  
    \centering  
    \includegraphics[width=0.8\textwidth]{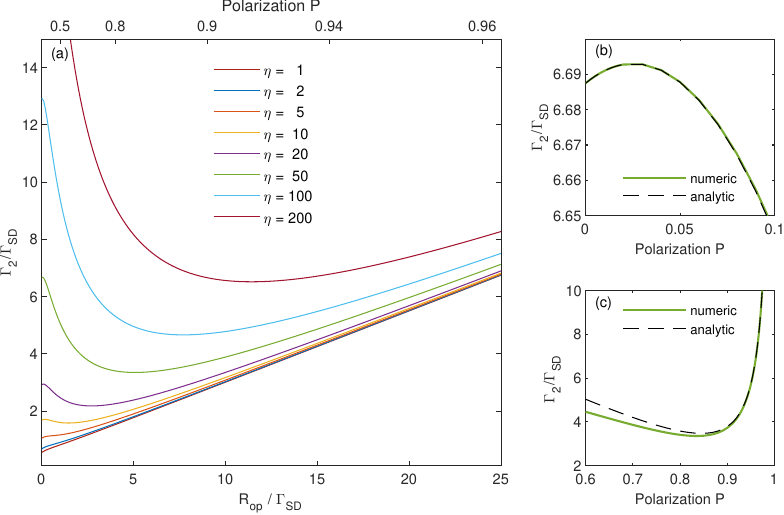}  
    \caption{(a) The linewidth of ${}^{87}{\rm Rb}$ atom as a function of spin polarization for different values of $\eta$. 
    (b) Zoom-in of the low-polarization part of the curve corresponding to $\eta =50$. The solid curve is the exact numerical result, and the dashed curve is the analytic result up to the second order correction of $P$ [see Eqs.~\eqref{eq:linewidth_0th_order_correction}-\eqref{eq:linewidth_2nd_order_correction}].
    (c) Zoom-in of the high-polarization part. The solid curve is the exact numerical result, and the dashed curve is the analytic result of Eq.~\eqref{eq:linewidth_high_Q}.}  
    \label{fig:linewidth}  
\end{figure*}

Assume the right- and left-eigenstates of the superoperator $\mathcal{G}_{\text{WDA}}$, sharing the same eigenvalue $i\Delta_1 + \lambda_j$, are denoted as $\left\vert \lambda_j\right)$ and $\left\{\lambda_j\right\vert$, respectively,
\begin{align}
    \mathcal{G}_{\text{WDA}} \left\vert \lambda_j\right) = \left(i\Delta_1 + \lambda_j\right) \left\vert \lambda_j\right),\\
    \left\{\lambda_j\right\vert \mathcal{G}_{\text{WDA}} = \left(i\Delta_1 + \lambda_j\right) \left\{\lambda_j\right\vert.
\end{align}
The observable is decomposed as 
\begin{align}
    \label{eq:observable}
    &S=\left( S_x\vert \tilde{\rho}_1\right) = \sum_j \left(S_x \big\vert \lambda_j\right)  \left\{\lambda_j\big\vert \tilde{\rho}_1\right) = - \frac{i\Omega_R}{2}\sum_j \frac{w_j}{i\Delta_j + \lambda_j }.
\end{align}
where the weight factor $w_j$ is
\begin{equation}
    w_j =\left\{\lambda_j \big\vert [S_x^{\copyright}]_{1, 0} \left[\mathcal{Q}_x\right]_{0, 1} \big\vert \lambda_j\right).
\end{equation}
As shown in Fig.~\ref{EigenvaluesOfG}, the normalized weight factors are actually dominated by the one corresponding to the smallest eigenvalue $\lambda_{\rm min}$ of $\mathcal{G}_{\text{WDA}}$.
The resonant line shape is very well approximated by a single Lorentizan function, whose linewidth is the real part of the smallest eigenvalue of $\mathcal{G}_{\text{WDA}}$.
In the following, we will focus on the smallest eigenvalue of $\mathcal{G}_{\text{WDA}}$, and study its behavior under different spin polarization and spin exchange rate.

\section{Results and Discussion}
Figure \ref{fig:linewidth}(a) illustrates the overall behavior of the ${}^{87}{\rm Rb}$ linewidth as a function of pumping rate and spin exchange rate. 
At low spin exchange rates (e.g., $\eta = 1$), the linewidth monotonically increases with pumping rate. 
In contrast, at high spin exchange rates ($ \eta> 5$), the linewidth initially decreases with increasing pumping rate before reaching a minimum value and then increases linearly with an increase in pumping rate.
This nonmonotonic behavior is the consequence of the competition between the narrowing effect due to the spin exchange at high spin polarization and the broadening induced by strong optical pumping.
Indeed, the light narrowing effect at high spin exchange rate was observed in the strong magnetic field cases, where the Zeeman resonances are well resolved \cite{bhaskarLightNarrowingMagnetic1981,appeltLightNarrowingRubidium1999}. 
To reveal the difference between the Zeeman resonances resolved and unresolved cases, we analyze the linewidth of alkali atoms in the low and high polarization limits, and compare with the result of previous studies.

\subsection{Low polarization limit ($P\ll 1$)}
To analyze the linewidth in the low polarization limit, the dimensionless matrix Eq.~\eqref{eq:dimensionless_GWDA} is expanded in the power series of $P$ up to the second order as
\begin{equation}
    \mathbb{G}_{\rm WDA}(\eta, P) = \mathbb{G}_{\rm WDA}^{(0)} + P \mathbb{G}_{\rm WDA}^{(1)} + P^2 \mathbb{G}_{\rm WDA}^{(2)},  \label{eq:GWDA_expansion}
\end{equation}
where
\begin{align}
\mathbb{G}_{\mathrm{WDA}}^{(0)}  =& \left( \eta + 1 \right)  \left[ \mathcal{A}_{\mathrm{SD}} \right]_{1, 1} -  \eta \left[ \mathcal{A}_{\mathrm{SE}}^{(\perp)} \right]_{1,1}^{(0)} , \label{eq:G0}\\
\mathbb{G}_{\mathrm{WDA}}^{(1)}  =& \left[ \mathcal{A}_{\mathrm{SD}} \right]_{1, 1} -  \frac{\eta+1}{2}  \left[ \mathcal{A}_{\mathrm{SE}}^{(z)} \right]_{1,1} - \eta \left[ \mathcal{A}_{\mathrm{SE}}^{(\perp)} \right]_{1,1}^{(1)},\label{eq:G1} \\
\mathbb{G}_{\mathrm{WDA}}^{(2)}  =& \left[ \mathcal{A}_{\mathrm{SD}} \right]_{1, 1}  -  \frac{1}{2}  \left[ \mathcal{A}_{\mathrm{SE}}^{(z)} \right]_{1,1}  -  \eta \left[ \mathcal{A}_{\mathrm{SE}}^{(\perp)} \right]_{1,1}^{(2)}.\label{eq:G2}
\end{align}
In Eqs.~\eqref{eq:G0}-\eqref{eq:G2}, we have expanded the matrix $\left[\mathcal{A}_{\rm SE}^{(\perp)}(P)\right]_{1, 1}$ to the second order of $P$, i.e., 
\begin{equation}
     \left[ \mathcal{A}_{\mathrm{SE}}^{(\perp)}(P) \right]_{1,1}  = \left[ \mathcal{A}_{\mathrm{SE}}^{(\perp)} \right]_{1,1}^{(0)}  + P \cdot \left[ \mathcal{A}_{\mathrm{SE}}^{(\perp)} \right]_{1,1}^{(1)}  + P^2 \cdot\left[ \mathcal{A}_{\mathrm{SE}}^{(\perp)} \right]_{1,1}^{(2)}.
\end{equation}

The leading order matrix $\mathbb{G}_{\rm WDA}^{(0)}$ is Hermitian, whose eigenvalue $\lambda_{k}^{(0)}$ and corresponding eigenvector $\left\vert v_{k}^{(0)}\right)$ are obtained by solving the eigenvalue problem
\begin{equation}
   \mathbb{G}_{\mathrm{WDA}}^{(0)} \left\vert v_{k}^{(0)} \right)  = \lambda_{k}^{(0)}\left\vert v_{k}^{(0)} \right) . 
\end{equation}
The smallest eigenvalue is denoted as $\lambda_0^{(0)}$. 
The correction to $\lambda_0^{(0)}$ due to the high-order matrices $\mathbb{G}_{\rm WDA}^{(1)}$ and $\mathbb{G}_{\rm WDA}^{(2)}$ is calculated using perturbation theory.
Up to the second order of $P$, the minimum eigenvalue is
\begin{equation}
    \lambda_0 = \lambda_0^{(0)} + \lambda_0^{(1)} P + \left(\lambda_{0}^{(2,a)} + \lambda_{0}^{(2,b)}\right) P^2 ,
\end{equation}
where
\begin{align}
\lambda_0^{(1)} &= \left( v_{0}^{(0)} \right\vert   \mathbb{G}_{\mathrm{WDA}}^{(1)}\left\vert v_{0}^{(0)}\right),  \\
\lambda_0^{(2a)} &=  \left( v_{0}^{(0)} \right\vert   \mathbb{G}_{\mathrm{WDA}}^{(2)}\left\vert v_{0}^{(0)} \right), \\
\lambda_0^{(2b)} &= \sum_{k\neq 0}\frac{  \left( v_{0}^{(0)} \right\vert \mathbb{G}_{\mathrm{WDA}}^{(1)} \left\vert v_{k}^{(0)} \right) \left( v_{k}^{(0)} \right\vert   \mathbb{G}_{\mathrm{WDA}}^{(1)} \left\vert v_{0}^{(0)} \right) }{\lambda_{0}^{(0)}-\lambda_{k}^{(0)}}.
\end{align}

For a given nuclear spin quantum number $I$, in the low polarization limit, the leading order contribution of linewidth is
\begin{align}
	\Gamma_{2}^{(0)} =  \lambda_0^{(0)} \Gamma_{\rm SD} =\frac{1}{q_{\rm SD}} \Gamma_{\rm SD} + \frac{1}{q_{\rm SE}} \Gamma_{\rm SE},\label{eq:linewidth_0th_order_correction}
\end{align}
where two slowing-down factors $q_{\rm SD}$ and $q_{\rm SE}$ are defined as
\begin{align}
    q_{\rm SD} &= \frac{(2I+1)^2}{2I^2+I+1}, \label{eq:slowing_down_sd}\\
    q_{\rm SE} &= \frac{3(2I+1)^2}{2I(2I-1)}. \label{eq:slowing_down_se}
\end{align}
The spin exchange slowing-down factor $q_{\rm SE}$ in Eq.~\eqref{eq:slowing_down_se} is identical to that obtained in the spin-exchange relaxation free (SERF) regime,
while the spin destruction slowing-down factor $q_{\rm SD}$ in Eq.~\eqref{eq:slowing_down_sd} is different from the longitudinal slowing-down factor at low polarization limit \cite{Seltzer2008} and that in the resolved Zeeman resonance regime (see Sect.~\ref{Sect:RZR}).

The first order correction of the linewidth with $P\ll 1$ is
\begin{equation}
	\Gamma_{2}^{(1)} = \lambda_0^{(1)}P\Gamma_{\rm SD}  = \frac{P}{q_{\rm SD}} \Gamma_{\mathrm{SD}} \approx  \frac{1}{q_{\rm SD}} R_{\rm op}.\label{eq:linewidth_1st_order_correction}
\end{equation}
Note that the first order correction is proportional to the optical pumping rate $R_{\rm op}$,
and it is independent on the spin-exchange rate $\Gamma_{\rm SE}$, which is different from the case of the RZR regime (see \ref{Sect:RZR}).
Furthermore, the correction in Eq.~\eqref{eq:linewidth_1st_order_correction} is always positive, which implies that the linewidth is always broadened by optical pumping when $R_{\rm op} \ll \Gamma_{\rm SD}$.

The light-narrowing effect manifests itself in the second order correction.
Unfortunately, the analytic expression of the second order correction for the general nuclear spin quantum number $I$ is lengthy. 
Here, we show the second order correction for $I=3/2$ (e.g., ${}^{87}{\rm Rb}$) as an example
\begin{equation}
   \Gamma_2^{(2)}  = \lambda_0^{(2)} P^2 \Gamma_{\rm SD} = \frac{7\Gamma_{\mathrm{SD}}^2 -34\Gamma_{\mathrm{SE}}\Gamma_{\mathrm{SD}}-98\Gamma_{\mathrm{SE}}^2}{560\Gamma_{\mathrm{SE}}\Gamma_{\mathrm{SD}} + 160\Gamma_{\mathrm{SD}}^2} P^2.\label{eq:linewidth_2nd_order_correction}
\end{equation}
The second order correction brings about a local maximum of the linewidth, as long as $\lambda_0^{(2)}<0$.
Figure~\ref{fig:linewidth}(b) shows the linewidth in the low polarization regime, with a direct comparison of the the numerical result and the perturbation analysis in Eqs.~\eqref{eq:linewidth_0th_order_correction}-\eqref{eq:linewidth_2nd_order_correction}.
Figure~\ref{fig:minmax}(a) presents the position of the local maximum at various spin exchange rate, obtained from the numerical calculation.
For the case of $I=3/2$, the local maximum exists when the ratio $\eta=\Gamma_{\rm SE}/\Gamma_{\rm SD}$ is above a threshold value $\eta_{0}=6.22$. 
This agrees with the analytic result in Eq.~\eqref{eq:linewidth_2nd_order_correction}, where $\lambda_0^{(2)}$ becomes positive if the spin exchange rate $\Gamma_{\rm SE}$ is too small.
For weak spin exchange rate (e.g., in low density vapor) with $\eta < \eta_0$, the local maximum does not exist, and the linewidth monotonically increases with the polarization as shown in Fig.~\ref{fig:linewidth}(a).

\subsection{High polarization limit ($Q=1-P\ll 1$)}

\begin{figure}[tbp]  
    \centering  
    \includegraphics[width=0.9\columnwidth]{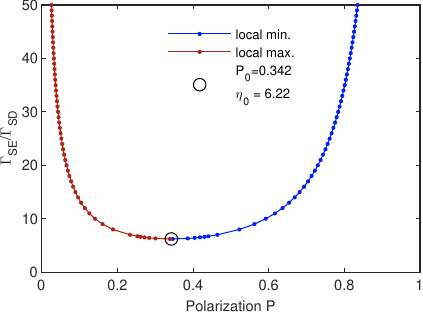}  
    \caption{Local minimum and local maximum of the linewidth of ${}^{87}{\rm Rb}$ as a function of spin polarization $P$.
    As increasing the spin-exchange rate, the position of the local maximum shifts to higher polarization, while the local minimum shifts in the opposite direction.
    The local minimum and local maximum merged together when the parameter $\eta = \eta_0$. 
    For $\eta < \eta_0$, the linewidth increases monotonically as increasing the polarization.}  
    \label{fig:minmax}  
\end{figure}

\begin{figure}[btp]  
    \centering  
    \includegraphics[width=0.9\columnwidth]{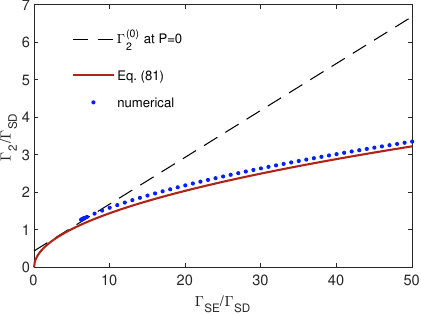}  
    \caption{Light narrowing effect. Blue dots are the numerical results of the local minimum of the linewidth as a function of $\eta$.
    For $\eta > \eta_0$, the linewidth of at the local minimum is smaller than that at $P=0$ (the dashed curve) due to the light-narrowing effect.
    The red dashed-dotted curve is the analytic result of Eq.~\eqref{eq:linewidth_local_minimum_analytic}.}  
    \label{fig:local_min}  
\end{figure}

Similar analysis is performed in the high polarization limit, where the matrix $\mathbb{G}_{\rm{WDA}}$ is expanded as
\begin{equation}
   \mathbb{G}_{\mathrm{WDA}}(\eta, 1-Q) = Q^{-1} \cdot \tilde{\mathbb{G}}_{\mathrm{WDA}}^{(-1)} + \tilde{\mathbb{G}}_{\mathrm{WDA}}^{(0)}  + Q\cdot \tilde{\mathbb{G}}_{\mathrm{WDA}}^{(1)},
\end{equation}
where
\begin{align}
\tilde{\mathbb{G}}_{\mathrm{WDA}}^{(-1)}  =& \left[ \mathcal{A}_{\mathrm{SD}} \right]_{1, 1} - \frac{1}{2} \left[ \mathcal{A}_{\mathrm{SE}}^{(z)} \right]_{1,1}, \label{eq:G_WDA_high_Q_0}\\
\tilde{\mathbb{G}}_{\mathrm{WDA}}^{(0)}  =& \eta   \left[ \mathcal{A}_{\mathrm{SD}} \right]_{1, 1} -  \frac{\eta -1}{2}  \left[ \mathcal{A}_{\mathrm{SE}}^{(z)} \right]_{1,1}  - \eta \left[ \tilde{\mathcal{A}}_{\mathrm{SE}}^{(\perp)} \right]_{1,1}^{(0)} ,\label{eq:G_WDA_high_Q_1} \\
\tilde{\mathbb{G}}_{\mathrm{WDA}}^{(1)}  =&  \frac{\eta}{2}  \left[ \mathcal{A}_{\mathrm{SE}}^{(z)} \right]_{1,1}  -  \eta \left[ \tilde{\mathcal{A}}_{\mathrm{SE}}^{(\perp)} \right]_{1,1}^{(1)}  .\label{eq:G_WDA_high_Q_2}
\end{align}
In Eqs.~\eqref{eq:G_WDA_high_Q_1} \& \eqref{eq:G_WDA_high_Q_2}, we have used the expansion of the transverse spin exchange matrix 
\begin{equation}
     \left[ \mathcal{A}_{\mathrm{SE}}^{(\perp)}(1-Q) \right]_{1,1}  = \left[ \tilde{\mathcal{A}}_{\mathrm{SE}}^{(\perp)} \right]_{1,1}^{(0)}  + Q \cdot \left[ \tilde{\mathcal{A}}_{\mathrm{SE}}^{(\perp)} \right]_{1,1}^{(1)} .
\end{equation}
Note that the matrix $\tilde{\mathbb{G}}_{\mathrm{WDA}}^{(-1)}$  is not Hermitian. 
With the eigenvalues $\tilde{\lambda}_k^{(-1)}$ and the corresponding right- and left-eigenvectors $\left \vert \tilde{v}_k^{(-1)}\right)$ 
and $\left \{ \tilde{v}_k^{(-1)}\right\vert$ of the leading order matrix $\tilde{\mathbb{G}}_{\mathrm{WDA}}^{(-1)}$, which satisfy the following equations
\begin{align}
    \tilde{\mathbb{G}}_{\mathrm{WDA}}^{(-1)} \left\vert\tilde v_{k}^{(-1)} \right)  &= \tilde \lambda_{k}^{(-1)}\left\vert \tilde v_{k}^{(-1)} \right),\\
    \left\{\tilde v_{k}^{(-1)} \right\vert\tilde{\mathbb{G}}_{\mathrm{WDA}}^{(-1)}   &= \tilde \lambda_{k}^{(-1)}\left\{ \tilde v_{k}^{(-1)} \right\vert,
\end{align}
the minimum eigenvalues expanded in a power series of $Q$ is
\begin{equation}
    \tilde{\lambda}_0= Q^{-1} \tilde{\lambda}_0^{(-1)} + \tilde{\lambda}_0^{(0)}  + \left(\tilde{\lambda}_0^{(1, a)} + \tilde{\lambda}_0^{(1, b)}\right) Q ,
\end{equation}
where
\begin{align}
\tilde{\lambda}_{0}^{(0)} &= \left\{ \tilde v_{0}^{(-1)} \right\vert   \tilde{\mathbb{G}}_{\mathrm{WDA}}^{(0)}\left\vert \tilde v_{0} ^{(-1)}\right), \label{eq:lambda_0_high_Q_0}\\
\tilde{\lambda}_{0}^{(1,a)} &= \left\{ \tilde v_{0}^{(-1)} \right\vert   \tilde{\mathbb{G}}_{\mathrm{WDA}}^{(1)}\left\vert \tilde v_{0} ^{(-1)}\right), \label{eq:lambda_0_high_Q_1}\\
\tilde{\lambda}_{0}^{(1,b)} &= \sum_{k\neq 0}\frac{  \left\{ \tilde v_{0}^{(-1)} \right\vert \tilde{\mathbb{G}}_{\mathrm{WDA}}^{(0)} \left\vert \tilde v_{k}^{(-1)} \right) \left\{ \tilde v_{k}^{(-1)} \right\vert   \tilde{\mathbb{G}}_{\mathrm{WDA}}^{(0)} \left\vert \tilde v_{0}^{(-1)} \right) }{\tilde \lambda_{0}^{(-1)}-\tilde \lambda_{k}^{(-1)}} \label{eq:lambda_0_high_Q_2}
\end{align}
With Eqs.~\eqref{eq:lambda_0_high_Q_0}-\eqref{eq:lambda_0_high_Q_2}, the linewidth in the high polarization limit is 
\begin{equation}
    \Gamma_2 = \tilde{\lambda}_0 \Gamma_{\rm SD} = \frac{1}{2I+1} R_{\rm op} + \frac{1}{2} \Gamma_{\rm SD} + \frac{1}{\xi_I} \frac{\Gamma_{\rm SD}\Gamma_{\rm SE}}{R_{\rm op}}.
    \label{eq:linewidth_high_Q}
\end{equation}
where $\xi_I$ is factor depending on the nuclear spin quantum number $I$.
The complete expressions of linewidth for different isotopes are listed in Table~\ref{table:linewidth_analytic}.
Although the coefficient in front of $\Gamma_{\rm SE}$ can be small in the high polarization limit with $R_{\rm op} \gg \Gamma_{\rm SD}$, the third term in Eq.~\eqref{eq:linewidth_high_Q} is not negligible if the spin exchange rate $\Gamma_{\rm SE}$ is large.

\setlength{\tabcolsep}{15pt}
\renewcommand{\arraystretch}{2.2}
\begin{table*}[btp]
\centering
\begin{tabular}{ccccc}
        \toprule
        $\displaystyle I$ & isotope & $\displaystyle \Gamma_2(P\to 0)$ & $\displaystyle \Gamma_2(P\to 1)$  & $\eta_0$ \\
        \hline
        $1$           & ${}^{6}{\rm Li}$ & $\displaystyle \Gamma_{2} =\frac{2}{27}\Gamma _{\text{SE}}+\frac{4}{9} \Gamma _{\text{SD}}$ & $\displaystyle \Gamma_2 =  \frac{R_{\mathrm{OP}}}{3} + \frac{\Gamma_{\mathrm{SD}}}{2} + \frac{\Gamma_{\rm SD}\Gamma_{\rm SE}}{8 R_{\rm op}}$ & 13.82\\
        $\displaystyle \frac{3}{2}$ & ${}^{7}{\rm Li}$, ${}^{23}{\rm Na}$, ${}^{39}{\rm K}$, ${}^{41}{\rm K}$, ${}^{87}{\rm Rb}$ & $\displaystyle \Gamma_{2} = \frac{1}{8}\Gamma _{\text{SE}}+\frac{7}{16} \Gamma _{\text{SD}}$ & $\displaystyle \Gamma_2 =  \frac{R_{\mathrm{OP}}}{4} + \frac{\Gamma_{\mathrm{SD}}}{2}+ \frac{5\Gamma_{\rm SD}\Gamma_{\rm SE}}{24 R_{\rm op}} $ & 6.22\\
        $\displaystyle \frac{5}{2}$ & ${}^{85}{\rm Rb}$ & $\displaystyle \Gamma_{2} =\frac{5}{27} \Gamma _{\text{SE}}+\frac{4 }{9}\Gamma _{\text{SD}}$ & $\displaystyle \Gamma_2 =  \frac{R_{\mathrm{OP}}}{6} + \frac{\Gamma_{\mathrm{SD}}}{2} + \frac{3\Gamma_{\rm SD}\Gamma_{\rm SE}}{10 R_{\rm op}}$ & 2.77\\
        $\displaystyle \frac{7}{2}$ & ${}^{133}{\rm Cs}$ & $\displaystyle \Gamma_{2}=\frac{7}{32} \Gamma _{\text{SE}}+\frac{29}{64} \Gamma _{\text{SD}}$ & $\displaystyle \Gamma_2 =  \frac{R_{\mathrm{OP}}}{8} + \frac{\Gamma_{\mathrm{SD}}}{2} + \frac{39\Gamma_{\rm SD}\Gamma_{\rm SE}}{112 R_{\rm op}}$ & 1.66\\
        \hline
\end{tabular}
\caption{Linewidth of different isotopes in low and high polarization limits. 
The analytic results of the linewidth $\Gamma_2$ are kept to the lowest order of polarization $P$ or $1-P$. 
See the text for higher order corrections.}
\label{table:linewidth_analytic}
\end{table*}

Equation~\eqref{eq:linewidth_high_Q} implies that the linewidth has a local minimum when changing the optical pumping rate $R_{\rm op}$.
Figure~\ref{fig:minmax} shows the position of the local minimum for different values of the spin exchange rate $\Gamma_{\rm SE}$. 
At high spin exchange rate (i.e., $\eta \gg 1$), the local minimum and local maximum are well separated.
As the spin exchange rate decreases, the local minimum and maximum of the linewidth are merged into a single point, denoted as $\left(\eta_0, P_0\right)$.
For weak spin exchange rate with $\eta < \eta_0$, the line width monotonically increases with increasing optical pumping rate $R_{\rm op}$.
To observe the light-narrowing effect, the spin exchange rate must be larger than the threshold value, that is, $\eta > \eta_0$.
The threshold values $\eta_0$ for different isotopes are listed in Table~\ref{table:linewidth_analytic}.

Figure~\ref{fig:local_min} compares the linewidth in the low polarization limit with the local minimum. 
Equation~\eqref{eq:linewidth_0th_order_correction} shows that the linewidth scales linearly with the spin-exchange rate at $P=0$,
while, according to Eq.~\eqref{eq:linewidth_high_Q}, the linewidth at the local minimum scales as $\sim \Gamma_{\rm SE}^{1/2}$, i.e.,
\begin{equation}
    \label{eq:linewidth_local_minimum_analytic}
    \Gamma_{2, {\rm min}} = 2\sqrt{\frac{\Gamma_{\rm SD}\Gamma_{\rm SE} }{(2I+1)\xi_I}},
\end{equation}
which agrees well with the numerical results (see Fig.~\ref{fig:local_min}).

\subsection{Comparison to the RZR regime}
\label{Sect:RZR}
\begin{figure}[bp]  
    \centering  
    \includegraphics[width=0.9\columnwidth]{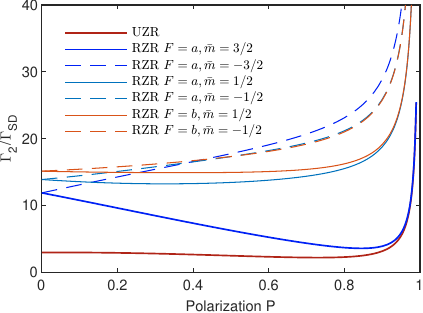}  
    \caption{Comparison the linewidth in the UZR and RZR regimes for the $I=3/2$ case. 
    In the low polarization limit, the linewidth in the RZR regime is larger than that in the UZR regime.
    In the high polarization limit, the two linewidths converge to approximately the same value.}  
    \label{fig:compare}  
\end{figure}

The linewidth in the RZR regime is \cite{Appelt1998, appeltLightNarrowingRubidium1999, jauIntenseNarrowAtomicClock2004}
\begin{align}
    \Gamma_{2, \bar{m}}^{\text{RZR}} &= \left( \Gamma_{\mathrm{SE}}  + \Gamma_{\mathrm{SD}} + R_{\mathrm{OP}}\right) \frac{3(2I+1)^2-4\bar{m}^2+1}{4(2I+1)^2}  \notag\\
    &- \frac{(P\Gamma_{\mathrm{SE}} + R_{\mathrm{OP}})\bar{m}}{2(F-I)(2I+1)} - \Gamma_{\mathrm{SE} } \frac{e^{\beta \bar{m}}}{Z_I} \frac{(2F+1)^2-4\bar{m}^2}{4(2I+1)^2},
    \label{eq:linewidth_RZR}
\end{align}
where $F=a$ or $F=b$.
Indeed, for a given Zeeman resonance labeled with $\bar{m}$, the linewidth in the RZR regime is the diagonal element of the matrix $\mathcal{G}_{\rm WDA}$ in Eq.~\eqref{FirstOrder1WDA}.
A direct comparison between the linewidth in the UZR and RZR regimes for the $I=3/2$ case is shown in Fig.~\ref{fig:compare}. Here, we focus on the narrowest Zeeman resonance, i.e., $\bar{m}= I$.

In the low polarization limit $P\ll 1$, the linewidth in the RZR regime is 
\begin{equation}
    \Gamma_{2, I}^{\text{RZR}} = \frac{\Gamma_{\mathrm{SD}}}{q_{\rm SD}'}    + \frac{ \Gamma_{\mathrm{SE}}}{q_{\rm SE}'} + \frac{I(2I+1)\Gamma_{\rm SD} -(4I+1)\Gamma_{\rm SE}}{(2I+1)^2} P,
\end{equation}
where the slowing-down factors are defined as
\begin{align}
    q_{\rm SD}' &= \frac{2I+1}{I+1},\label{eq:slowing_down_sd1} \\
    q_{\rm SE}' &= \frac{(2I+1)^2}{I(2I+3)} . \label{eq:slowing_down_se1}
\end{align}
The smaller slowing-down factors in Eq.~\eqref{eq:slowing_down_sd1} \& \eqref{eq:slowing_down_se1} compared to Eq.~\eqref{eq:slowing_down_sd} \& \eqref{eq:slowing_down_se} 
causes larger linewidth in the RZR regime than in the UZR regime.
Furthermore, the sign of the coefficient in front of $P$ depends on the ratio $\eta = \Gamma_{\rm SE}/\Gamma_{\rm SD}$.
As increasing $\Gamma_{\rm SE}$, the coefficient become negative and the light-narrowing effect occurs when 
\begin{equation}
    \frac{\Gamma_{\rm SE}}{\Gamma_{\rm SD}} > \frac{I(2I+1)}{4I+1}. \label{eq:eta_condition}
\end{equation}
This is different from the UZR regime, where the first order correction in Eq.~\eqref{eq:linewidth_1st_order_correction} is always positive, 
and the second order correction in Eq.~\eqref{eq:linewidth_2nd_order_correction} drives the light-narrowing effect.

In the high polarization limit, the linewidth in the RZR regime is
\begin{equation}
    \Gamma_{2, I}^{\rm RZR} = \frac{1}{2I+1}R_{\mathrm{OP}} + \frac{I+1}{2I+1} \Gamma_{\mathrm{SD}} + \frac{I}{2I+1} \frac{\Gamma_{\mathrm{SD}}\Gamma_{\mathrm{SE}}}{R_{\mathrm{OP}}}.
    \label{eq:linewidth_highP_RZR}
\end{equation}
As the linewidth is dominated by strong optical pumping, the linewidth in the RZR regime [see Eq.~\eqref{eq:linewidth_highP_RZR}] is approximately the same as that in the UZR regime [see Eq.~\eqref{eq:linewidth_high_Q}], as shown in Fig.~\ref{fig:compare}.
Physically, this result is reasonable because, in the high polarization limit, the Zeeman resonances with $\bar{m} < I$ are less populated, and the interplay between different Zeeman resonances becomes less important.

\section{Conclusions}
We studied the linewidth of the magnetic resonance of alkali-metal atoms in the UZR regime.
A theoretical framework is developed to describe the dynamics of the Zeeman coherences and the populations, in presence of the optical pumping, spin exchange, and spin destruction processes.
Under the RWA and WDA, the master equation is solved, and important phenomena such as the light-narrowing effect and the light-broadening effect are discussed.
We obtain the linewidth of alkali metal atoms with different nuclear spin $I$, as functions of the spin polarization parameter $P$ and the relative spin-exchange strength $\eta$.
Analytic analysis of the linewidth in the UZR regime is provided in the low- and high-polarization limits.
Based on the analytic results, a direct connection between the microscopic rates (e.g. the spin exchange rate $\Gamma_{\rm SE}$ and the spin destruction rate $\Gamma_{\rm SD}$) and the experimentally observable resonance linewidth is established.
We show that the linewidth in the UZR regime is different from previous results in the RZR regime, due to the interplay between different Zeeman resonances.
Our study provides a deep understanding and useful theoretical tools for developing various atomic sensors (e.g. atomic magnetometers or comagnetometers) working in the UZR regime.


%

\end{document}